\documentclass{elsart}
\journal{New Astronomy}

\usepackage{natbib}

\usepackage{graphicx}
\usepackage{epsfig}

\usepackage{amssymb}
\newcommand{\kmsmpc}{\kms\;{\rm Mpc}^{-1}}
\newcommand{\lya}{Ly$\alpha$\ }
\newcommand{\hkpc}{h^{-1}{\rm kpc}}
\newcommand{\hmpc}{h^{-1}{\rm Mpc}}

\newcommand{\kms}{\;{\rm km}\,{\rm s}^{-1}}

\begin{document}

\begin{frontmatter}



 \title{Simulations of Early Galaxy Formation}


\author{Romeel Dav\'e$^1$}
\address{$^1$ Department of Astronomy, University of Arizona, Tucson, AZ 85721\\
E-mail: rad@as.arizona.edu}

\begin{abstract}
We present the predictions for the photometric and emission line
properties of galaxies present during the latter stages of reionization
from $z=8\rightarrow 6$.  These preliminary predictions are made from
cosmological hydrodynamic simulations that include star formation and
feedback, but not the effects of radiative transfer.  We find significant
numbers of galaxies that have stellar masses exceeding $10^8 M_\odot$
by $z=8$, with metallicities in the range of one-tenth solar.  These
galaxies are just beyond the reach of current near-infrared surveys, but
should be found in large numbers by next-generation programs.  The \lya\
luminosity function does not evolve much from $z=6$ to $z=8$, meaning that
it should also be possible to detect these objects in significant numbers
with upcoming narrow band surveys, unless the escape fraction of \lya\
evolves significantly between those epochs.
\end{abstract}


\end{frontmatter}

\section{Introduction}

A key goal of next generation ground and space-based telescopes is to
detect the objects responsible for reionizing the universe at $z>6$
\cite{fan02}.  Few such objects have been observed up till now, and
hence their detailed physical properties remain poorly constrained.
For optimal design and usage of future telescopes and instruments,
it is important to understand the nature of these early galaxies.

Simulations of galaxy formation can provide insights into the expected
properties of these systems within the context of a well-established
framework for hierarchical structure formation.  While the properties
of dark matter halo assembly are relatively well understood owing to
precisely determined cosmological parameters \cite{spe03}, the properties
of the gas clouds that condense to form observable stars remain subject
to a large range of theoretical uncertainty.

It is possible that galaxy formation in the pre-reionization universe
proceeds much differently than today or in the recent past, owing to
high densities, short cooling times, low metallicities, and a flat
power spectrum on mildly nonlinear scales.  Furthermore, the process of
reionization itself may have large effects on the galaxy population,
either in terms of suppressing star formation and stimulating it, as is
detailed in various contributions to these proceedings.  In short, it
is possible that our recipes and models for galaxy formation tuned to
today's observations may have little relevance at these early epochs.
However, this view is debatable.  It is worth remembering that galaxies
that are most likely to be observable in the near future are among the
largest and/or most vigorously star forming galaxies at those epochs,
and that such galaxies are unlikely to be forming their very first
generation of stars.  Furthermore, these systems may be relatively
unaffected by the global reionization process occuring in the universe
at large, because by virtue of their high bias they reside in a portion
of the universe that reionizes earlier.

In these proceedings we study the observable and physical properties of
reionization epoch galaxies in cosmological hydrodynamic simulations of
galaxy and structure formation.  We include a subset of relevant physical
processes such as kinetic feedback with accompanying metal injection
and metal-line cooling, in addition to the usual processes required
for galaxy formation.  These models are known to work reasonably well in
predicting the post-reionization galaxy population \cite{nig05,fin05}, but
it remains unclear if they are truly applicable in the pre-reionization
universe.  Hence this work is preliminary in that it does not include the
effects of radiative transfer that may be an important driver for how
star formation proceeds in these systems.  We also do not explicitly
account for possible variations in the stellar initial mass function
due to the presence of low-metallicity stars.  Nevertheless, we will
show that the objects that are likely to be actually observable at these
epochs tend to be the ones for which such reionization epoch processes
are expected to be comparatively less relevant.

\section{Simulations}
\label{sec:sims}

We employ the parallel N-body+SPH code Gadget-2 \cite{spr02} in this
study.  This code uses an entropy-conservative formulation of smoothed
particle hydrodynamics along with a tree-particle-mesh code for handling
gravity.  It includes the effects of radiative cooling assuming ionization
equilibrium, which we have extended to include metal-line cooling based
on Sutherland \& Dopita \cite{sut93}.  The metal cooling function
is interpolated to the gas metallicity as tracked self-consistently
by Gadget-2.  Stars are formed using a recipe that reproduces the
Kennicutt relation \cite{ken98}, using a subgrid multi-phase model that
tracks condensation and evaporation in the interstellar medium following
\cite{mck77}.  Stars inherit the metallicity of the parent gas particle,
and from then on cannot be further enriched.

Gadget-2 has an observationally-motivated prescription for driving
superwinds out of star forming galaxies, which is tuned to produce
agreement with the global star formation history of the universe
\cite{spr03}.  In \cite{dav05} we describe modifications on this scheme
designed to better match certain observations; in this work, we will
only employ two superwind schemes: A ``constant wind" model where all
the particles entering into superwinds are expelled at 484~km/s out of
star forming regions and a constant mass loading factor of 2 is assumed
\cite{spr03}, and a ``momentum wind" model where the imparted velocity
is proportional to the local velocity dispersion (computed from the
potential) and the mass loading factor is inversely proportional to
the velocity dispersion, as expected in momentum-driven wind scenarios
\cite{mar05,mur05}.  Using larger simulations evolved to low redshift,
we have verified that both models broadly reproduce the star formation
history of the universe.  These two models are meant to illustrate the
sensitivity of our results to the somewhat ad hoc feedback prescription.

Gadget-2 implements heating due to a photoionizing background, taken from
the 2001 CUBA model \cite{haa01} which assumes a 10\% escape fraction
of ionizing radiation from galaxies in addition to the contribution from
observed quasars, and results in reionization occurring at $z\approx 9$.
Spatial uniformity is assumed, which is obviously incorrect in detail
during reionization, but may nevertheless be a passable approximation
if the galaxies of interest reionize their surroundings earlier and
their masses are significantly above that which would be suppressed
by photoionization.

Our two simulations with different superwind models each have a comoving
volume of $8\hmpc$, with an equivalent Plummer softening of $0.6\hkpc$
(i.e. better than 100~pc physical at $z>6$).  A WMAP-concordant
cosmology is assumed, with $\Omega=0.3$, $\Lambda=0.7$, $H_0=70\kmsmpc$,
$\sigma_8=0.9$, and $\Omega_b=0.04$.  These models were run with $256^3$
dark matter and $256^3$ gas particles, from $z=249$ to $z=6$.  The mass
resolution is $4.8\times 10^5 M_\odot$ for gas particles and $3.1\times
10^6 M_\odot$ for dark matter.  Given the canonical value of 64 particles
for robust identification of halos, this means we are resolving halos
above $2\times 10^8 M_\odot$, which means that primordial atomic gas
should be the dominant coolant in all resolved systems.  Note that we
do not include H$_2$ cooling.

We identify galaxies using Spline Kernel Interpolative DENMAX
\cite{ker05}, and consider only galaxies with stellar masses exceeding
64 star particles \cite{fin05}.  We use Bruzual \& Charlot population synthesis models
\cite{bru03} to convert the simulated star formation history of each
galaxy into broad band colors in a variety of infrared filters, including
$J$, $K_s$, and the Spitzer $[3.6\mu]$ and $[24\mu]$ bands (quoted
in AB magnitudes).  We have examined other bands as well but all the
relevant trends can be gleaned from this set.  We account for dust using
a prescription based on the galaxy's metallicity and a locally-calibrated
metallicity-extinction relation, as described in \cite{fin05}.  We also
predict \lya emission line properties based on the instantaneous star
formation rates in each galaxy as output by Gadget-2; we describe this
in more detail later.

\section{Masses \& Metallicities}
\label{sec:phys}

\begin{figure}[!t]
\centerline{ \psfig{file=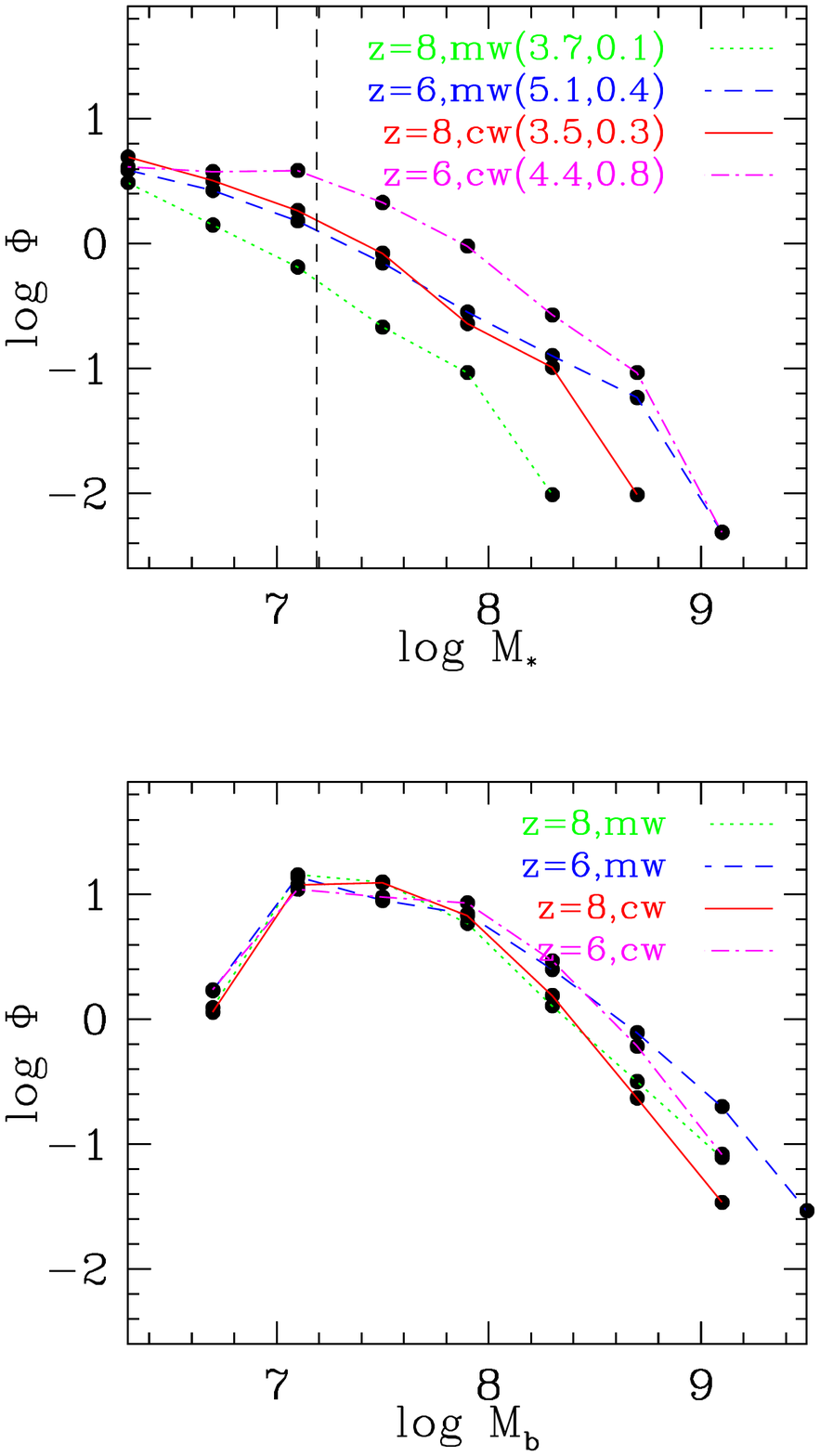, width=3.5in, angle=0}
\psfig{file=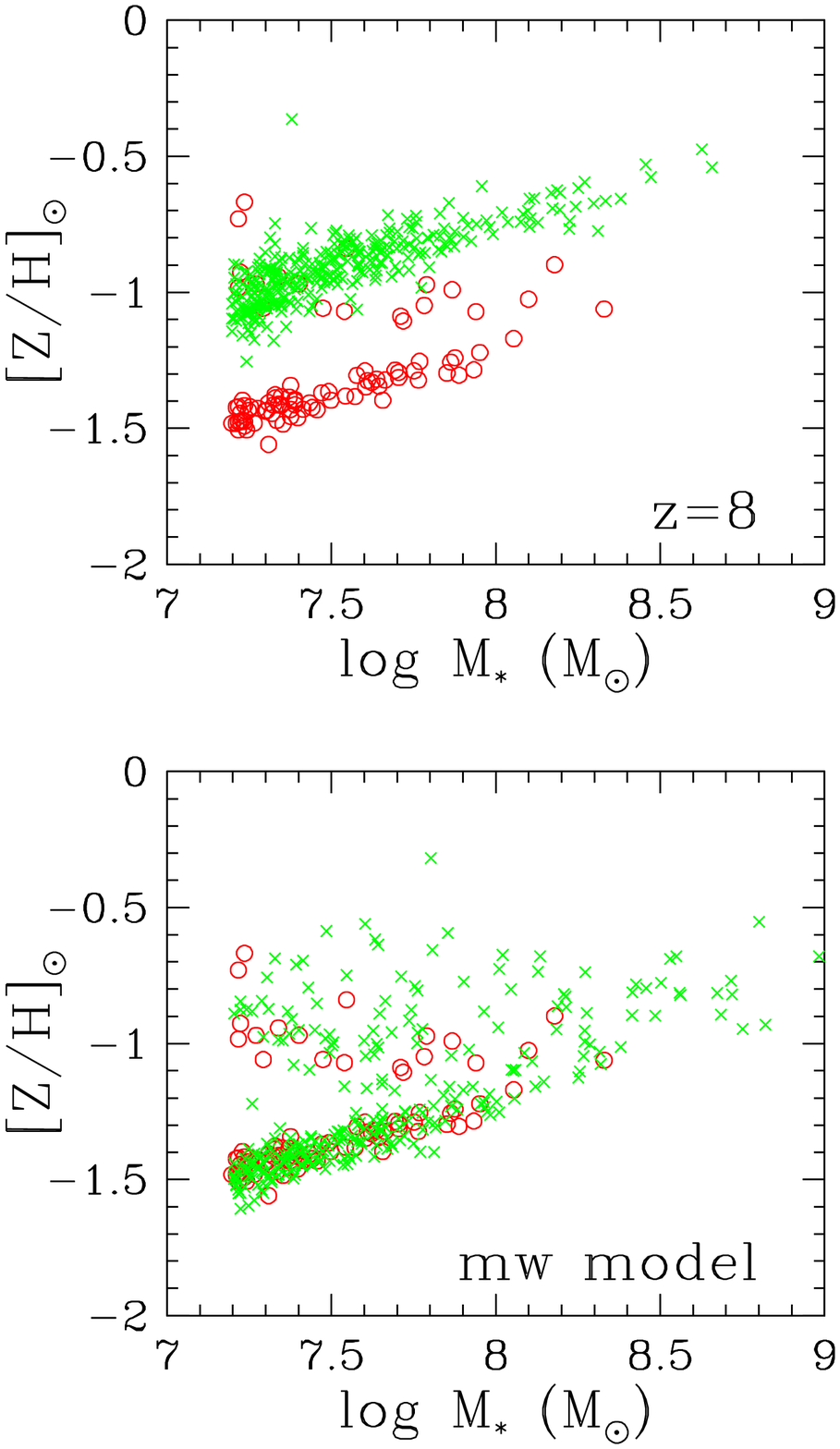, width=3.5in, angle=0} }
\caption{{\it Top Left:} Stellar mass functions of galaxies at $z=8$ and $z=6$ 
in the momentum wind model (dotted green and dashed blue lines, 
respectively), and constant wind model (solid red and dot-dashed
magenta lines).  Units are number per cubic $\hmpc$ per log stellar mass.
Vertical dashed line is our galaxy stellar mass resolution limit of 64 
star particles.
The numbers in parantheses in the captions represent the percentage of 
baryons in galaxies and in stars, respectively, for each model.
{\it Bottom Left:} Same as above panel, for total baryonic mass (cold 
dense gas + stars) functions.
{\it Top Right:} Gas-phase metallicity of galaxies at $z=8$
versus stellar mass, for the constant and momentum wind models (green 
and red points, respectively).  
{\it Bottom Right:} Metallicity versus stellar mass at $z=8$ (red circles)
and $z=6$ (green crosses) in the momentum wind model.
}
\label{fig:mfcomp}
\end{figure}

Here we show some basic physical properties of reionization-epoch galaxies
in our simulations.  Mass functions at $z=8$ and $z=6$ are shown in the
left panels of Figure~\ref{fig:mfcomp}, for the two superwind models.
In either model, a significant number of galaxies with stellar masses
exceeding $10^8 M_\odot$ are already in place by $z=8$, even within
our rather small $8\hmpc$ box.  By $z=6$, there are several galaxies
exceeding $10^9 M_\odot$.  The discovery of a $10^9 M_\odot$ system at
$z\sim 7$ by \cite{ega05} is therefore not unexpected.  Note that the
superwind feedback prescription has a significant effect on the stellar
mass function, producing nearly half a dex reduction in the mass function
for the momentum wind model as compared with the constant wind model.
Nevertheless, at the bright end the different wind models converge,
as might be expected since the winds will have a smaller effect on
larger systems.  There is less of a difference due to winds for the
total baryonic mass function (bottom left panel).

Metal pollution has a large effect on the reionization epoch, both
in terms of the cooling curve that strongly affects the Jeans mass in
star forming regions (see e.g. Schneider, these proceedings), as well as
enabling the formation of dust that can strongly affect the detectability
of systems, particularly in the \lya\ emission line.  In the upper right
panel of Figure~\ref{fig:mfcomp} we show the gas-phase metallicity of
galaxies in the constant (green crosses) and momentum (red circles)
wind models at $z=8$, and the lower right panel shows metallicities in
the momentum wind model at $z=6$ (green crosses) and $z=8$ (red circles).
In all cases, the metallicities of the galaxies exceed about one-thirtieth
solar.  While small, this is still far above the metallicity purported
to enact the transition to a Population II stellar population from
a Population III one \cite{bro04}.  The stellar metallicities are
similar to the gas, typically lower by less than a factor of two.
The feedback mechanism has a significant effect on the metallicity,
as might be expected because feedback carries metals out of galaxies.
There is little redshift evolution in the mass-metallicity relation,
with the exception of an increase in low-mass metal rich galaxies
that have been polluted by the winds from nearby larger systems.
While feedback processes are poorly constrained even at $z=0$ and could
therefore operate in a completely different fashion than assumed in our
simulations at these epochs, it is nevertheless interesting that massive
galaxies at these epoch already have metallicities that approach one-tenth
solar or more.  Indeed, by $z=4$ these same massive systems grow to have
near solar metallicities \cite{fin05}, in agreement with observations.
Hence at face value, these high redshift systems should already have
little if any contribution from Population III stars, and may contain a
significant amount of dust depending on the timescale for dust formation.

\section{Observable Properties}
\label{sec:observable}

\begin{figure}[!t]
\centerline{ \psfig{file=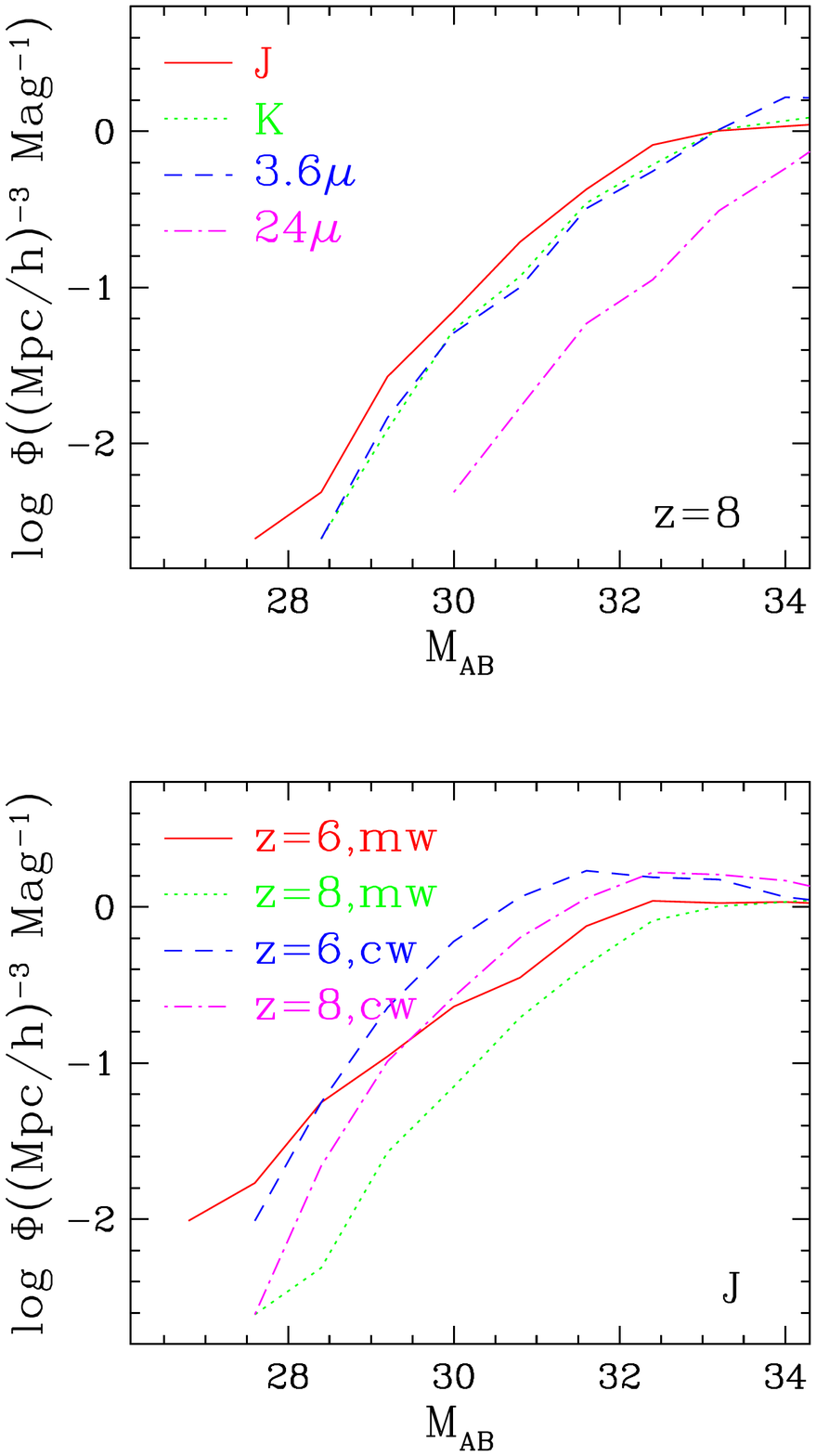, width=2.9in, angle=0}
\psfig{file=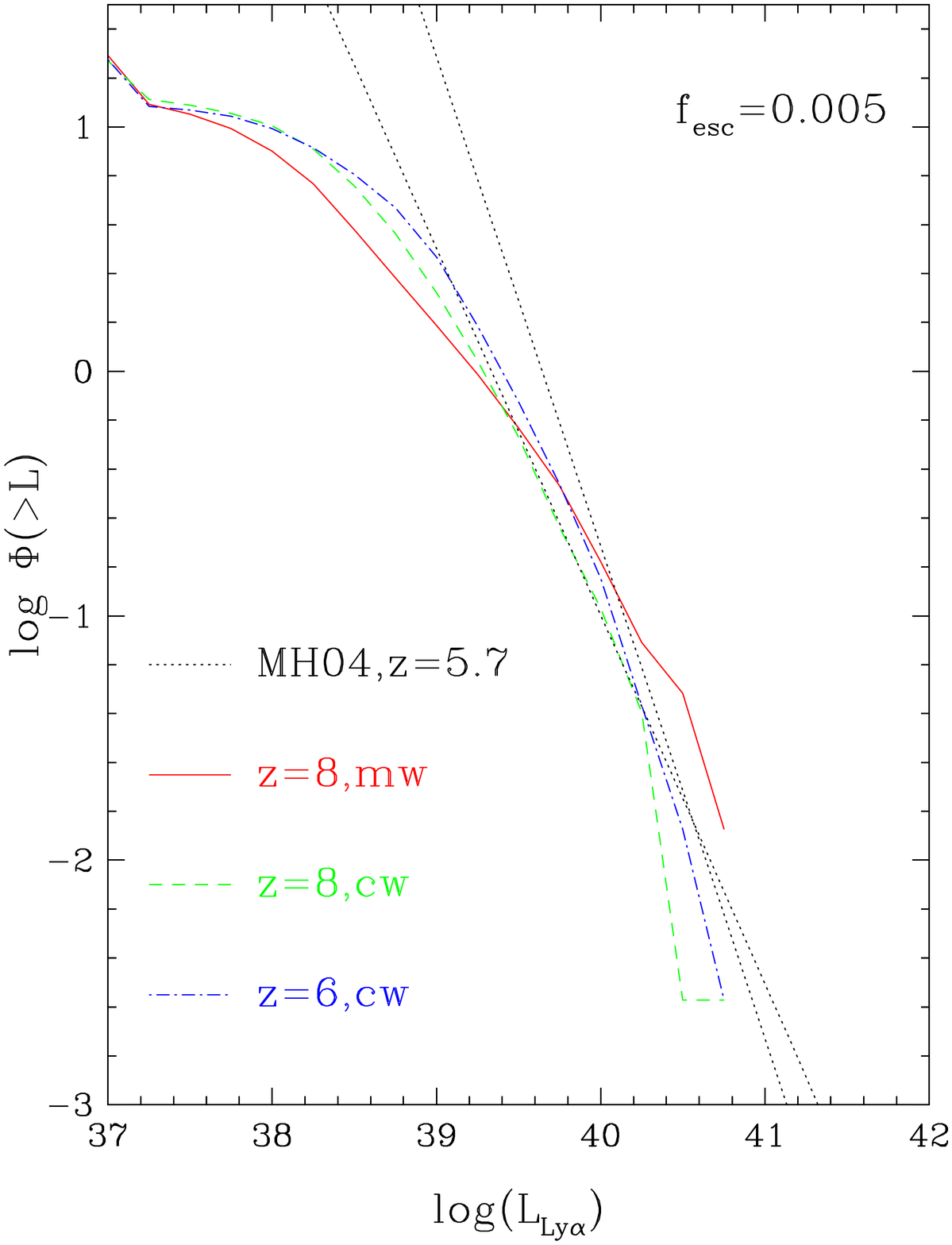, width=2.8in, angle=0} }
\caption{{\it Top Left:} Luminosity functions of galaxies in the
``constant wind" models for various observed bands.  {\it Bottom
Left:} Luminosity function evolution from $z=8\rightarrow 6$ in the
``constant wind" (cw) and ``momentum wind" (mw) models.  {\it Right:}
The Lyman alpha cumulative luminosity function evolution at $z=8$ and 6
in the constant wind model, with a Ly$\alpha$ escape fraction of 0.5\%
(i.e. the fraction of the stellar Ly$\alpha$ output that is observed).
A fit to data compiled by \cite{mal04} (assuming a faint-end slope of
-1.5) is extrapolated to faint luminosities and shown for comparison; the
escape fraction of 0.5\% was chosen to match the data for the constant
wind model.  The $z=8$ luminosity function for the momentum wind model
is also shown for comparison.
}
\label{fig:lumfcnz}
\end{figure}

We now examine some basic observable properties of these high-redshift systems.
We present luminosity functions (LFs) of simulated galaxies in the left
panels of Figure~\ref{fig:lumfcnz}.  The top panel shows the $z=8$
LF in various bands, for the momentum wind (mw) model.  The bands
$J\rightarrow [24\mu]$ correspond approximately to rest-frame 1500\AA\
to K-band.  As might be expected for these young galaxies, they are quite
blue, with the rest-UV being typically one magnitude brighter than rest-K.

The bottom panel shows the LF in the $J$ band for our two superwind models
at two redshifts: $z=8$ and $z=6$.  The momentum wind model suppresses
low-mass galaxy formation more effectively, and so the majority of the
evolution is in the bright end.  The constant wind model, conversely,
has a steep faint-end slope due to the comparatively low mass loading
factor in small galaxies, and shows mostly luminosity evolution between
the two redshifts.  The turnover at faint magnitudes is due to our
64 star particles resolution limit for galaxy stellar masses.

From these LFs we predict that a future $z\sim 8$ Lyman break survey
using a $z$-dropout technique and selecting down to 30th magnitude in $J$
would find $\sim 100$ galaxies per square arcminute per unit $\Delta z$,
plus or minus half a dex depending on model assumptions.  Even down
to 29th magnitude such a survey would find a few of galaxies per square
arcminute per $\Delta z$.  Note that due to our small simulation volume,
we are probably underestimating the bright end of the LF, but these values
are unlikely to increase by more than a factor of a couple.  These numbers
suggest that detecting significant samples of reionization-epoch galaxies
using dropout techniques is not too far removed from current technology.

Turning to \lya\ emission galaxies, it may be possible to detect such
systems using narrow-band surveys, and indeed such surveys are underway
(see various contributions in these proceedings, including Smith,
Willis, Jensen, and Stark).  As discussed in \cite{bar04}, there are large
uncertainties regarding the detectability of \lya\ emission from star
formation. Scattering due to the surrounding IGM including the damping
wing may make \lya\ emission more diffuse and difficult to detect.  As we
showed in \S\ref{sec:phys}, significant metals are expected to be present
in the most rapid star formers, so dust extinction may be considerable.
On the other hand, if the initial mass function is more top-heavy in
the early universe owing to the lower metallicities, then perhaps \lya\
emission is boosted \cite{sch03}.

In Figure~\ref{fig:lumfcnz} (right panel), we present a comparison of
the \lya\ luminosity function to observations compiled by \cite{mal04}.
Their data is best fit by a faint-end slope that is somewhere between
$-1.5$ and $-2$; those best fits are shown as the dashed lines.  To obtain
\lya\ luminosity functions, we can convert the instantaneous star
formation rate of a galaxy as given by Gadget-2 into a \lya\ luminosity
using a factor given by the models of \cite{sch03}:  For 1/20th solar
(typical for our galaxies) and a Salpeter IMF, the \lya\ luminosity
is $4.01\times 10^{42}$~erg/s/$M_\odot$.  Of course, not all the \lya\
emitted by young stars will actually reach us; there is an escape fraction
(defined here as the fraction of \lya\ photons we can observe, not the
fraction that escapes from the star forming region or galaxy itself) that
is highly uncertain.  We constrain this escape fraction by comparing our
$z=6$ predicted \lya\ luminosity function with the observations shown.
As is evident from  Figure~\ref{fig:lumfcnz}, an escape fraction of 0.5\%
produces an amplitude that is broadly in agreement with observations.

If we assume that this escape fraction does not evolve with redshift,
then the predictions for the $z=8$ luminosity functions are as shown.
With this assumption, there is relatively little evolution predicted
from $z=8\rightarrow 6$, though the bright end does evolve somewhat.
However, the \lya\ escape fraction may evolve rapidly either up or down
with redshift at these epochs \cite{bar04}, making these predictions
uncertain.  As there have been several \lya\ emitters seen at $z\approx
6.5$, it seems unlikely that all \lya\ emitters are made undetectable
by neutral hydrogen in the IGM.  However, beyond this, it is difficult
to make concrete statements about the expected observability of \lya\
emitters at $z>7$.  The results of observational programs such as those
described at this meeting as well as DAzLE \cite{hor04} will be critical
for constraining models.

\section{Discussion}
\label{sec:discussion}

We have presented results for the physical and observable properties of
reionization epoch galaxies from cosmological hydrodynamic simulations.
Our simulations show that high-redshift galaxies can already contain
sufficient stellar mass to be in agreement with the available
observations; masses of $10^8 M_\odot$ and above are common by $z\sim 8$.
The large masses also imply significant metal enrichment in these systems;
while they are far below solar and comparable to the lowest metallicity
dwarfs today such as I~Zw~18, their metallicities are still well above the
putative threshold for ushering in a more normal IMF as compared with
an extremely top-heavy IMF proposed for Population III stars.  Hence it
may be unwise to expect significant populations of near metal-free stars
in these systems.

These galaxies may be observable using broad-band selection techniques
or via their \lya\ emission.  The broad-band properties are probably
a more robust prediction of these models, and suggest that surveys
down to 30th magnitude (AB) in $J$ or $K$, or even [3.6$\mu$], would
yield a dense sampling of at least tens of objects per square arcminute.
The numbers may be higher by up to $\sim 1$~dex depending on which of
our feedback models turns out to be closer to reality.  Predictions of
\lya\ emitters are less robust because of large uncertainties in \lya\
transfer out of these systems, but if their properties are similar to
\lya\ emitters observed at $z\approx 5.7$, then surveys will have to
below $10^{40}$~ergs/s over sizeable areas to achieve a significant
sample of objects.

The outlook for detecting these systems is promising.  Current NICMOS
data can achieve depths of 28.5 in near-IR bands \cite{bou05}, albeit
in small fields.  In addition to JWST, several future space-based
telescopes have been proposed that will do large-area near infrared
surveys, including some versions of JDEM (Joint Dark Energy Mission),
as well as the proposed Galaxy Evolution Origins Probe (R. Thompson, PI).
Though these missions remain firmly in the distant and uncertain future,
the scientific richness of deep wide NIR surveys ensures that they are
likely to be accomplished at some point.  On the \lya\ emitter side, DAzLE
will be the premier instrument for such surveys in the immediate future,
but given their sensitivity it would be surprising if they found more than
a handful of $z>7$ systems.  Narrow band filters in NIR night sky windows
on future 20--30m class telescopes seems the best option for achieving
significant samples of these objects \cite{bar04}.  The timescales for
such facilities is of the same order as the space-based NIR ones, so it
will be an interesting race to see which technique matures first.

The author thanks B. Oppenheimer and K. Finlator for assistance in running
and analyzing the simulations, and V. Springel and L. Hernquist for providing
us with Gadget-2.

\end{document}